\documentclass[twocolumn,preprintnumbers,amsmath,amssymb,showpacs,prb]{revtex4-1}

\usepackage{graphicx}
\usepackage{dcolumn}
\usepackage{bm}

\begin{document}

\preprint{}

\title{Cavity mode identification for coherent terahertz emission\\
from a nearly square stack of intrinsic Josephson junctions} 

\author{M. Tsujimoto}%
 \email{tsujimoto@sk.kuee.kyoto-u.ac.jp}
\author{I. Kakeya}
\affiliation{%
Department of Electronic Science and Engineering, Kyoto University, Kyoto,
Nishikyo-ku, Kyoto 615-8510, Japan
}%

\author{T. Kashiwagi}
\author{H. Minami}
\author{K. Kadowaki}
\affiliation{%
Division of Materials Science, Faculty of Pure and Applied Sciences, University of Tsukuba, Tennodai 1-1-1, Tsukuba, Ibaraki 305-8573, Japan
}%

\date{\today}

\begin{abstract} 
Stacks of intrinsic Josephson junctions in Bi$_2$Sr$_2$CaCu$_2$O$_{8+\delta }$ emit intense and coherent terahertz waves determined by the internal electromagnetic cavity resonance.  We identify the excited transverse magnetic mode by observing the broadly tunable emissions from an identical nearly square stack and simulating the scattering spectrum.  We employ a wedge-type interferometer to measure emitted integral power independently of the far-field pattern.  The simulation results are in good agreement with observed resonance behaviors as a function of frequency.
\end{abstract}

\pacs{74.50.+r, 74.72.-h, 85.25.Cp}
\maketitle


\section{INTRODUCTION}
Electromagnetic (EM) waves in the 0.3--10~THz frequency range are thought to have great potential in research and industry~\cite{Kawase04}; thus, compact, solid-state, and continuous wave (CW) terahertz sources have been developed in the field of semiconductors and lasers~\cite{Tonouchi07}.  Presently, quantum cascade lasers, which emit radiation at frequencies greater than 1.45~THz and must be cooled to 37~K~\cite{Kumar11}, are of particular interest.  Room-temperature terahertz sources based on resonant tunneling diodes can generate frequencies up to 1.31~THz~\cite{Kanaya12} but emission power greater than 1~mW is presently not possible.

Intense, continuous, and broadly-tunable terahertz generation from high transition temperature (high-$T_c$) superconductor Bi$_2$Sr$_2$CaCu$_2$O$_{8+\delta }$ (Bi-2212) has attracted the attention of experimenters and theoreticians~\cite{Welp13}.  DC voltage $V_0$ applied to the stack of intrinsic Josephson junctions (IJJs)~\cite{Kleiner92} in single crystal Bi-2212 leads to AC current and concomitant terahertz emission at the Josephson frequency~\cite{Josephson62} 
\begin{equation}
f_{J} = \frac{2e}{h} \cdot \frac{V_{0}}{N},
\end{equation}
where $e$, $h$, and $N$ are the elementary charge, Planck's constant, and the number of active IJJs, respectively.

To generate intense terahertz waves, several tens of micron-sized mesa structures are milled out of Bi-2212.  According to the electromagnetism for the microstrip antenna~\cite{Carver81}, the transverse magnetic (TM) mode is excited at the resonance frequency: in the case of
a rectangular mesa, 
\begin{equation}
f_{mp}^c = \frac{c_0}{2 \sqrt{\epsilon}} \sqrt{\left( \frac{m}{w} \right) ^2 + \left( \frac{p}{\ell } \right) ^2 },  
\end{equation}
where $c_0$, $\epsilon$, $w$, and $\ell $ are the speed of light in vacuum, the dielectric constant, and the width and length of the mesa, respectively.    The emitted far-field waves are polarized with their E-field parallel to the plane that contains the wave vector of the standing EM wave in the mesa cavity.  The two integers ($m,p$) in Eq.~(2) correspond to the number of line nodes of the TM ($m,p$) mode standing wave.  In most cases using long rectangular mesas, the TM ($1,0$) mode with formation of the standing wave along the width of the mesa at $f^c_{1,0} = c_0 / 2\sqrt{\epsilon}w$ has been observed~\cite{Ozyuzer07,Kadowaki10}.  The similar collective resonance-like behavior along the length was found in laser microscopy~\cite{Wang10}, which, however, is not direct evidence for the cavity resonance.  To our knowledge, there still remains a lack of consensus on the issue on the cavity resonance condition.  Moreover, broadly tunable emissions obscure the role of the internal cavity resonance~\cite{Tsujimoto12-01,Kitamura14}.  To correctly identify the excited modes, the emitted integral wave has to be collected using lenses or mirrors with small f number because the far-field emission patterns are affected by the excited mode~\cite{Kashiwagi11}.

A complete study of the cavity resonance will help to better understand the mechanisms of coherent emission from the IJJ stack, which will make possible to control the polarization of the emitted wave and may facilitate the design of CW terahertz sources for coherent communication applications.  In this study, we show that various cavity modes with different ($m,p$) can be excited using an identical stack of IJJs by changing the bias condition.  We simulate the experimental data using the method of moments to calculate the resonance frequency and visualize the EM standing wave in the emitting mesa.

\section{EXPERIMENTAL SETUP}
We show experimental results from a nearly square mesa sample with $w = 100$~$\mu $m, $\ell = 140$~$\mu $m, and $t = 1.3$~$\mu $m.  The sample preparation method is described in Ref.~\cite{Tsujimoto12-01}.  All sample dimensions are measured using an atomic force microscope.  The horizontal dimensions ($w$, $\ell$) are averages at the top and bottom lengths.  Figure 1 shows the top view of experimental setup.  The sample was mounted on a He-flow cryostat.  The bath temperature $T_b$ was monitored using a thermometer attached to the sample holder.  For measuring the current-voltage characteristics (IVC), we used a conventional electrical circuit for measuring the resistance, where a load resistance of 100~$\Omega $, a standard resistor of 10~$\Omega $, and the emitting sample were connected in series to the constant voltage source.  During the $I$-$V$ measurements, we also monitored the far-infrared emission power using a silicon composite bolometer.  The bolometric signal was calibrated separately using a sub-millimeter power meter.  For the lock-in detection, the emitted EM waves were optically chopped at 80~Hz.  To collect the emitted integral waves, we used hemispherical silicon and plastic lenses and an off-axis parabolic mirror, as shown in Fig.~1.  The collimation effect owing to the silicon lens is described in detail in the Appendix A.  Note that the estimated solid angle of the detection window of 0.6~sr is more than 20 times larger than that in previous experiments; therefore, we can directly measure the total emission power from the sample.  At the focal point of the parabolic mirror, we set a wedge-type interferometer system to measure the emission frequency $f$~\cite{Tsujimoto12-02}.  The details of the frequency measurement are described in the Appendix B.

\begin{figure}[t] 
	\includegraphics[width=0.47\textwidth ,clip]{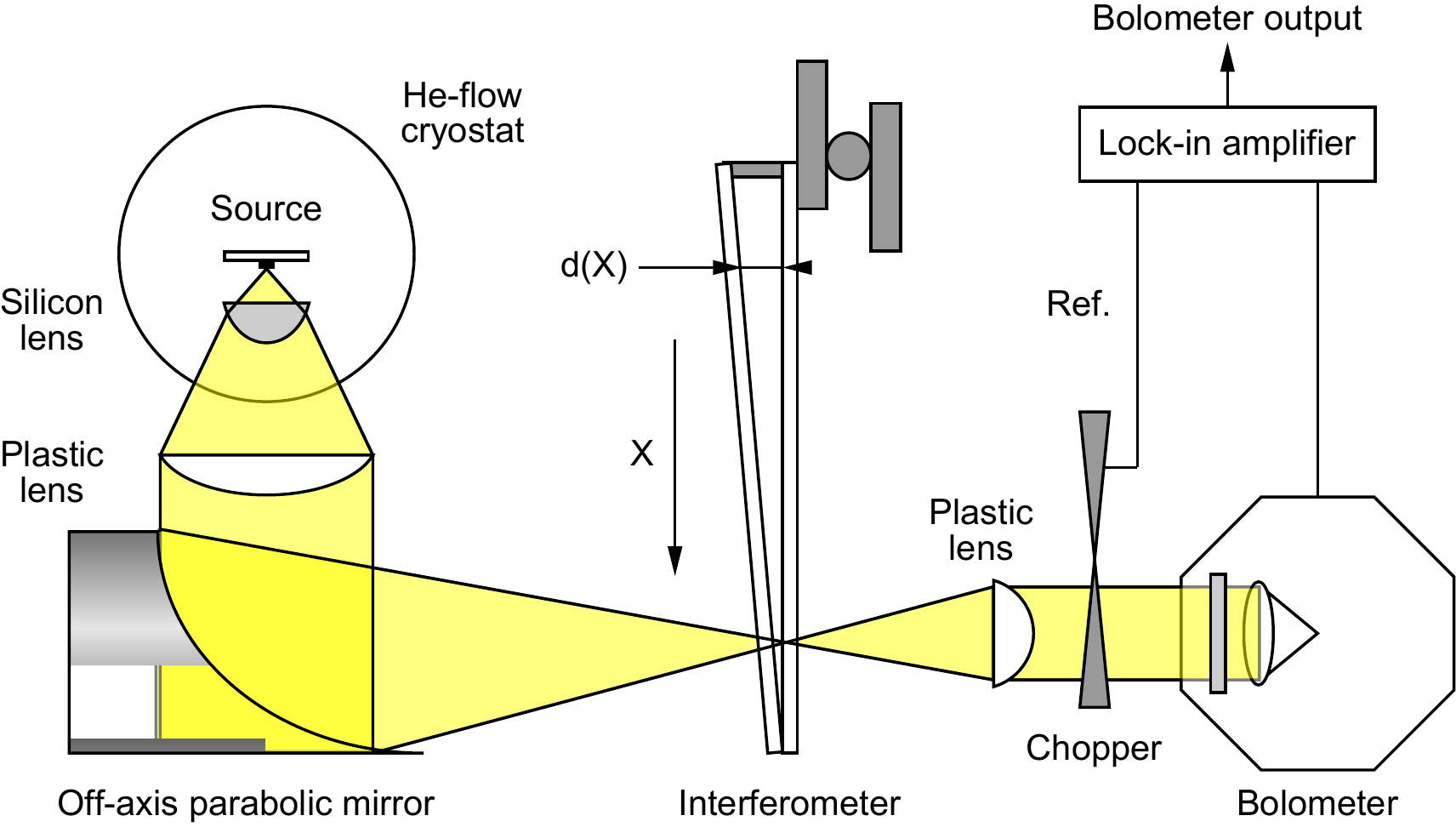} \\
	\caption{
Top view of the collection optical system.
}
\end{figure}%

\section{RESULTS AND DISCUSSION}
Figure 2(a) shows the IVC curve at $T_b=30$~K obtained using a cyclic bias scan.  We subtracted the contact resistance owing to the two terminal measurements from the $I$-$V$ data.  In Fig.~2(b), the emission power on the outermost IVC curve, which was simultaneously monitored using the bolometer, is shown as a function of the current.  The inset in Fig.~2(b) shows a scanning ion microscopy image of the emitting sample.  Four emission peaks are observed at 22.1, 15.0, 10.2, and 6.54~mA, as indicated by the arrows.  Each emission peak corresponds to the excitation of the different cavity mode and is discussed elsewhere.  We scanned the bias current upward and downward several times to confirm the emission reproducibility and found that the emission peaks always appeared at the same bias points.

\begin{figure}[t] 
	\includegraphics[width=0.47\textwidth ,clip]{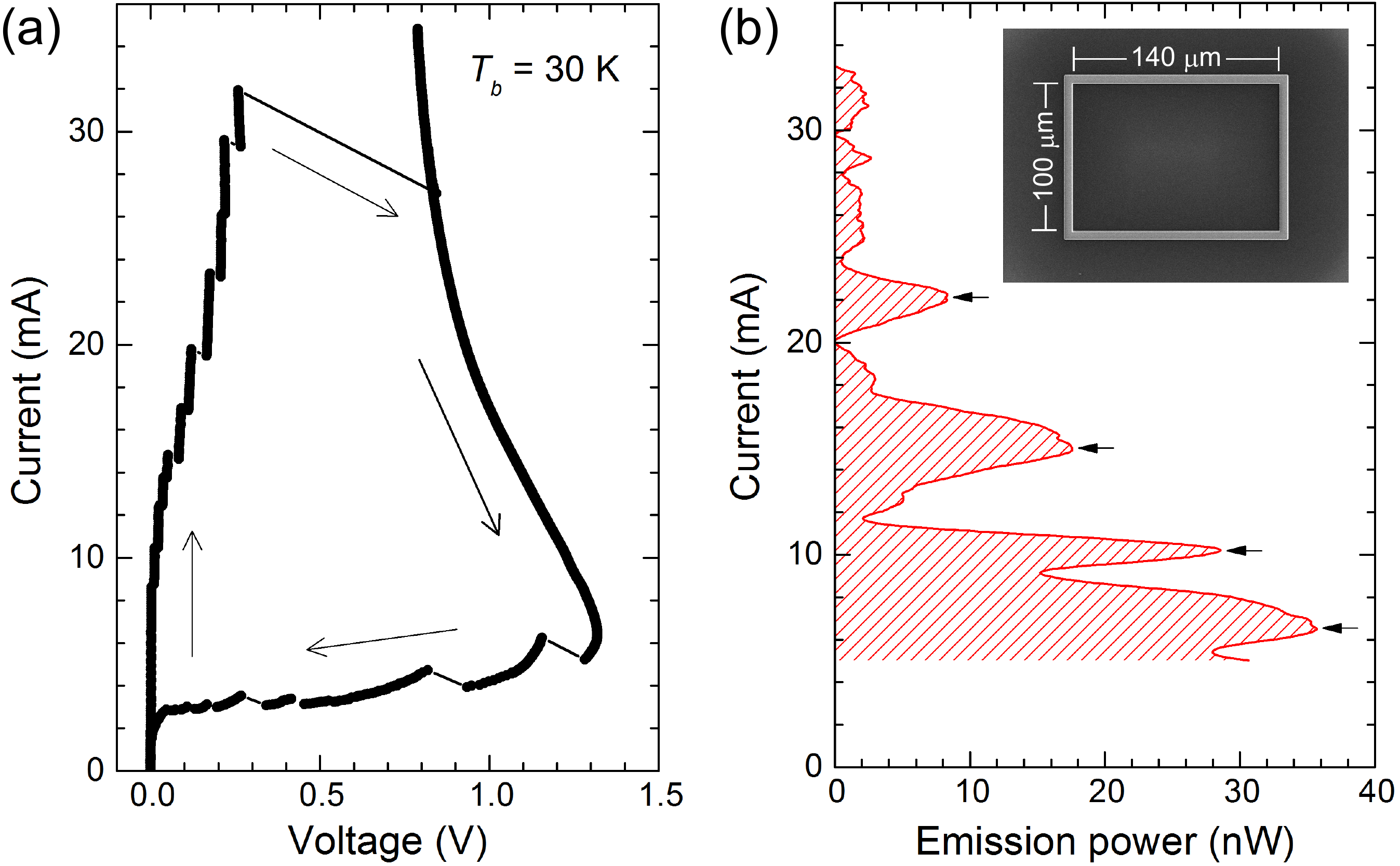} \\
	\caption{
(a) Current-voltage characteristics at $T_b=30$~K, obtained by the cyclic bias scan.  
(b) Emission power of the outermost IVC curve measured using the bolometer as a function of $I$.  
The inset shows a scanning ion microscopy image of the emitting sample.
}
\end{figure}%

The emission frequencies $f(V)$ at $T_b=30$~K are shown in Fig.~3 together with the emission power.  The error bars for $f$ are standard deviations from the sinusoidal fitting in the interferometer analysis.  The emission power in the right axis is the same data as in Fig.~2(b). Three bias points at 0.92, 1.11, and 1.26~V correspond to the emission peaks at TM ($0,2$), ($1,2$), and ($2,0$), respectively.  Note that $f$ for the rightmost peak of TM ($2,1$) cannot be obtained properly because the bias point around 1.35~V is less stable to the bias scan.  The dashed line represents the Josephson relation $f = f_{J} = (2e/h) V/N$, where $N = 851$ is obtained from the least squares fit to the experimental data.  This $N$-value is consistent with the value of 850 that is separately obtained from the height of the stack.

According to the Josephson relation, the bias voltage can be directly converted into the emission frequency.  For instance, the application of 1.00 V voltage to the $N=851$ IJJ stack corresponds to a frequency of 0.568~THz.   Hence, the plot of the emission power versus $V$ can be assumed to be equivalent to the resonance spectrum.  This assumption is supported by the fact that the observed $f$ corresponds exactly to $f_{J}$ within instrumental error (see Fig.~3).  Most importantly, because we make appropriate use of the collecting lenses, the behavior of the bolometer signal is commensurate with the emitted integral power although we neglect the light loss owing to the Fresnel reflections at the lens surfaces.

Figure 4 shows the emission power from the outermost IVC curve versus $V$ at $T_b=15$, 20, 25, and 30~K at different offsets.  The inset in Fig.~4 shows the $T_b$ dependence of the outermost IVC curve.  With increasing $T_b$, the IVC hysteresis and the concomitant maximum voltage are suppressed because of the reduction in the quasiparticle tunnel resistance.  The voltage condition for the TM ($m,p$) mode can be obtained using Eqs.~(1) and (2) with $f_{J}=f_{mp}^c$
\begin{equation}
V_{mp}^c = \frac{Nh}{2e} \cdot f_{mp}^c = \frac{Nh}{2e} \cdot \frac{c_0}{2 \sqrt{\epsilon}} \sqrt{\left( \frac{m}{w} \right) ^2 + \left( \frac{p}{\ell } \right) ^2 }.  
\end{equation}
The six vertical lines represent the calculated $V_{mp}^c$ for ($0,2$), ($1,2$), ($2,0$), ($2,1$), ($1,3$), and ($2,2$), respectively.  Note that in the calculation of $V_{mp}^c$, we use only the realistic sample dimensions ($w$, $\ell$) and $\epsilon =17.6$~\cite{Kadowaki10}.  We observe that the emission power has local maximum at each $V^c_{m,p}$ and its peak height varies depending on ($m,p$).  The slight difference between calculated $V^c_{m,p}$ and observed peaks of emission power arises from the typical 5--10\% difference in the top and bottom lengths of the actual stack due to the trapezoidal cross section.  Particularly, the TM ($2,0$) and TM ($2,1$) modes have relatively high peaks in these modes.  This observed behavior is consistent with the simulation results, which is discussed later.  Note that the TM ($1,0$) mode cannot be identified in Fig.~4, since the resonance voltage of $V_{10}^c=0.631$~V falls short of the applicable voltage range.

\begin{figure}[t] 
	\includegraphics[width=0.42\textwidth ,clip]{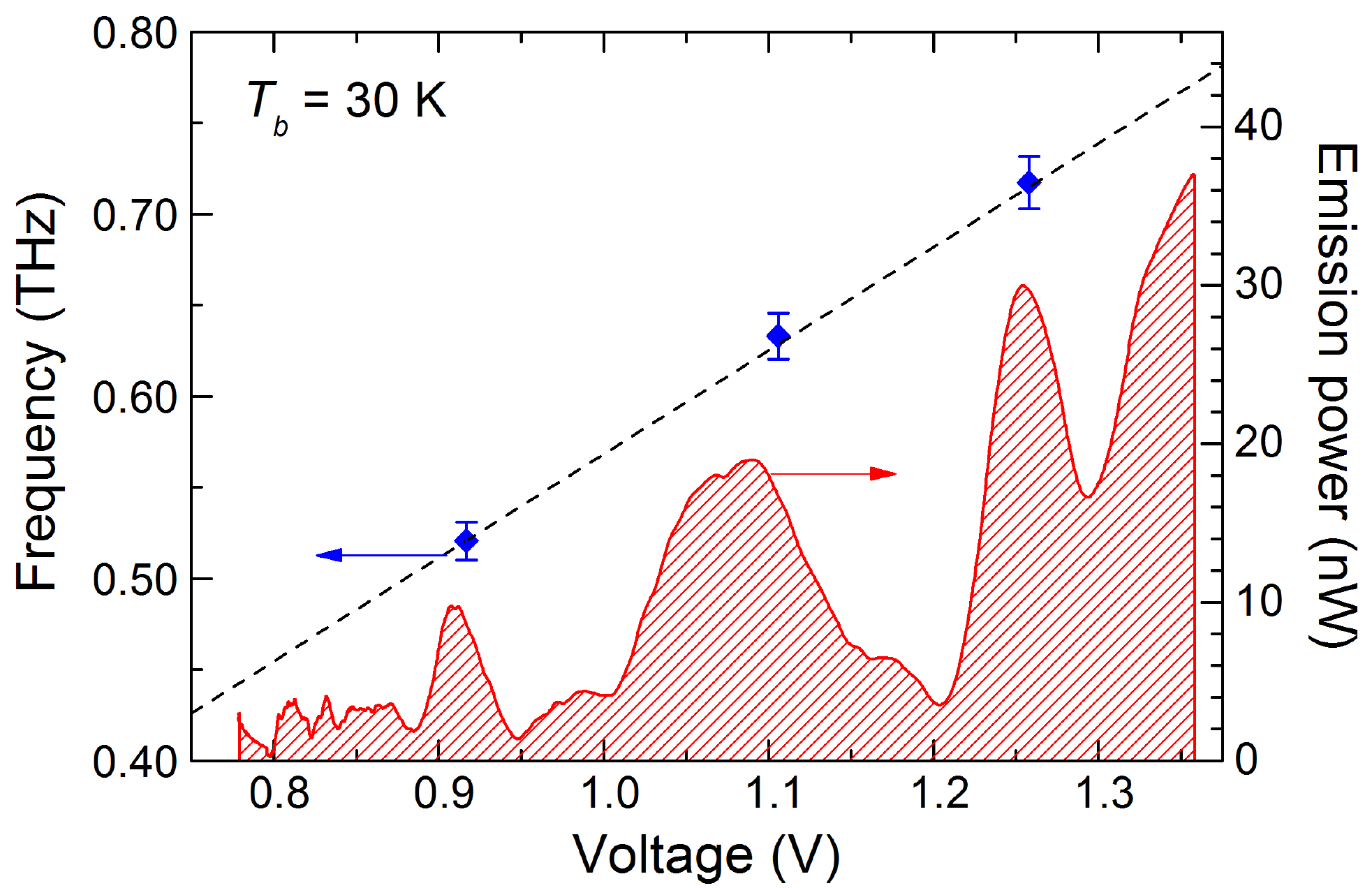} \\
	\caption{
Emission frequency measured using the interferometer system (left) and the detected emission power (right) versus voltage.  The dashed line represents the Josephson relation $f = f_{J} = (2e/h) V/N$ for $N = 851$.
}
\end{figure}%

Each emission peak in Fig.~4 has a nearly symmetric width of $\Delta V_{mp}^c$ that may reflect the quality of the cavity resonance.  None of the obtained $\Delta V_{mp}^c$ values is smaller than the calculation error of $V_{mp}^c$ that stems from difference in the top and bottom length of the actual stack.  An effective $Q$-value can be estimated from $Q_{mp}^c = V_{mp}^c / \Delta V_{mp}^c$;  for example, $Q_{20}^c= 27$ at $T_b = 30$~K.   The observed low-$Q$ cavity resonance suggests that the broadening effect of the emission is independent of the angle of the sidewall inclination in the mesa but is caused by the impedance mismatching.  Moreover, a mixing experiment showed a 23~MHz spectral linewidth at the minimum~\cite{Li12}.  With careful consideration of the minimum resolution limit of 0.75~GHz of the FT-IR spectrometer~\cite{Kakeya13}, a $Q$-value in the order of 10$^3$--10$^4$ is estimated, which is much larger than $Q_{mp}^c$.  This high-$Q$ resonance is interpreted as a unique feature of the synchronized IJJ stack; however, further synchronization studies are needed to provide supporting information.

\begin{figure}[t] 
	\includegraphics[width=0.47\textwidth ,clip]{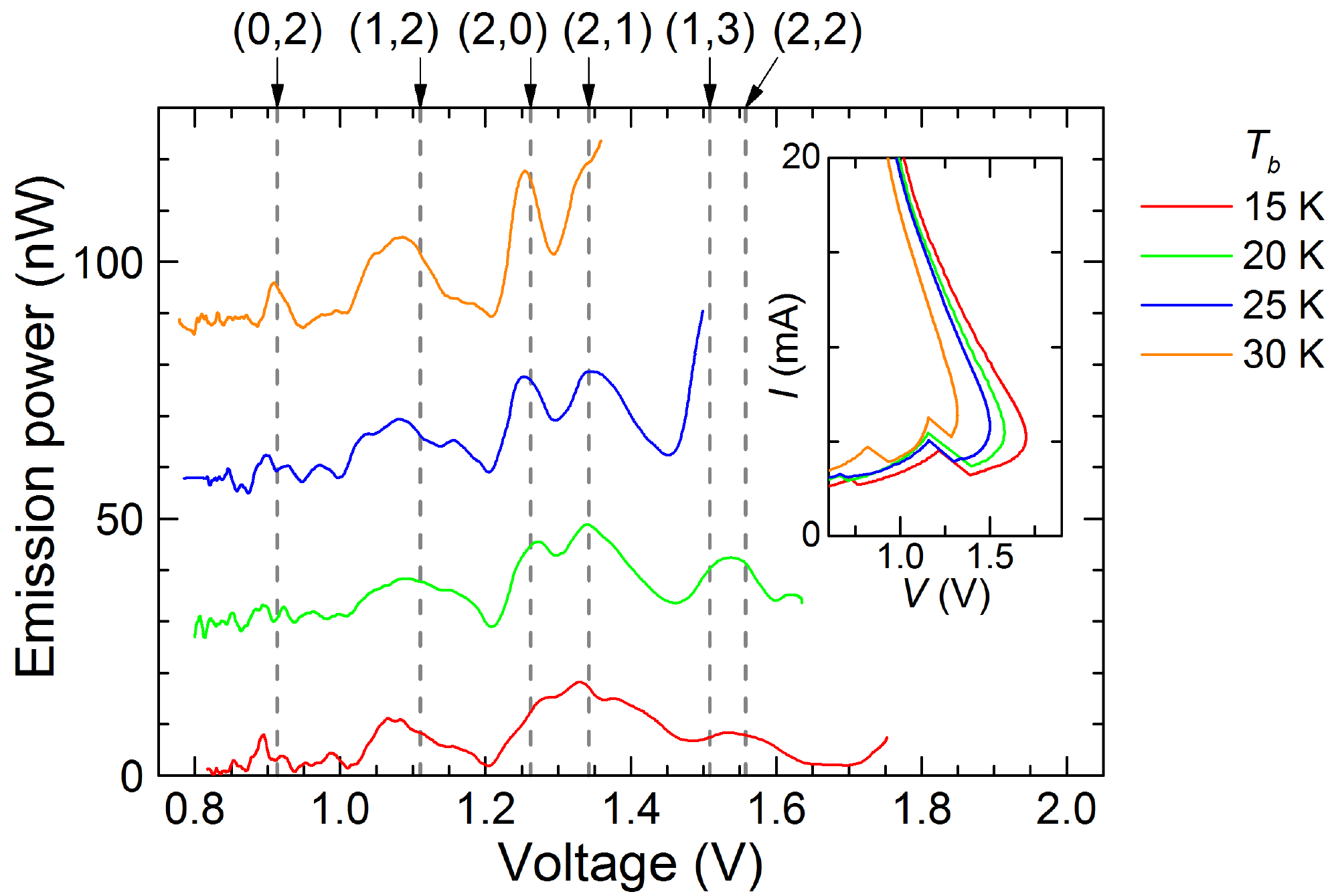} \\
	\caption{
Observed emission power from the outermost IVC curve at different offsets versus voltage at various $T_b$.  Dashed lines represent the calculated $V_{mp}^c$ using Eq.~(3) for the TM ($m,p$) modes.  The inset shows the $T_b$ dependence of the outermost IVC curve.
}
\end{figure}%

The Joule heating over the entire mesa comes off the backbending IVC curve in the resistive state.  In the high-bias regime, the temperature distribution in the emitting stack is inhomogeneous due to the enormous local heating~\cite{Tsujimoto14}.  This would lead to a shorten length of the internal cavity~\cite{Wang10}.  Nevertheless, the overall peak structure in Fig.~4 is insensitive to $T_b$, which implies that the effective cavity condition including $w$, $\ell$, and $\epsilon $ is nearly independent of $T_b$ in the measurement range.

\section{SIMULATION}
We use the commercial simulator Sonnet~\cite{Sonnet} to analyze the experimental result in terms of the cavity condition.  By using the method of moments with Sonnet, we can calculate the EM field from a numerical integral of Green's function at the discretized surface of the patch.  The current density at the patch surface can also be calculated by substituting the EM field into Maxwell's equations.  Although the simulation results hereinafter presented do not allow us to take into account the synchronization among the numerous IJJs, which is actually very important for the physics of coherent emission, we find that they are consistent with the experimental results and thus informative to readers.

 \begin{figure*}[t] 
	\includegraphics[width=0.78\textwidth ,clip]{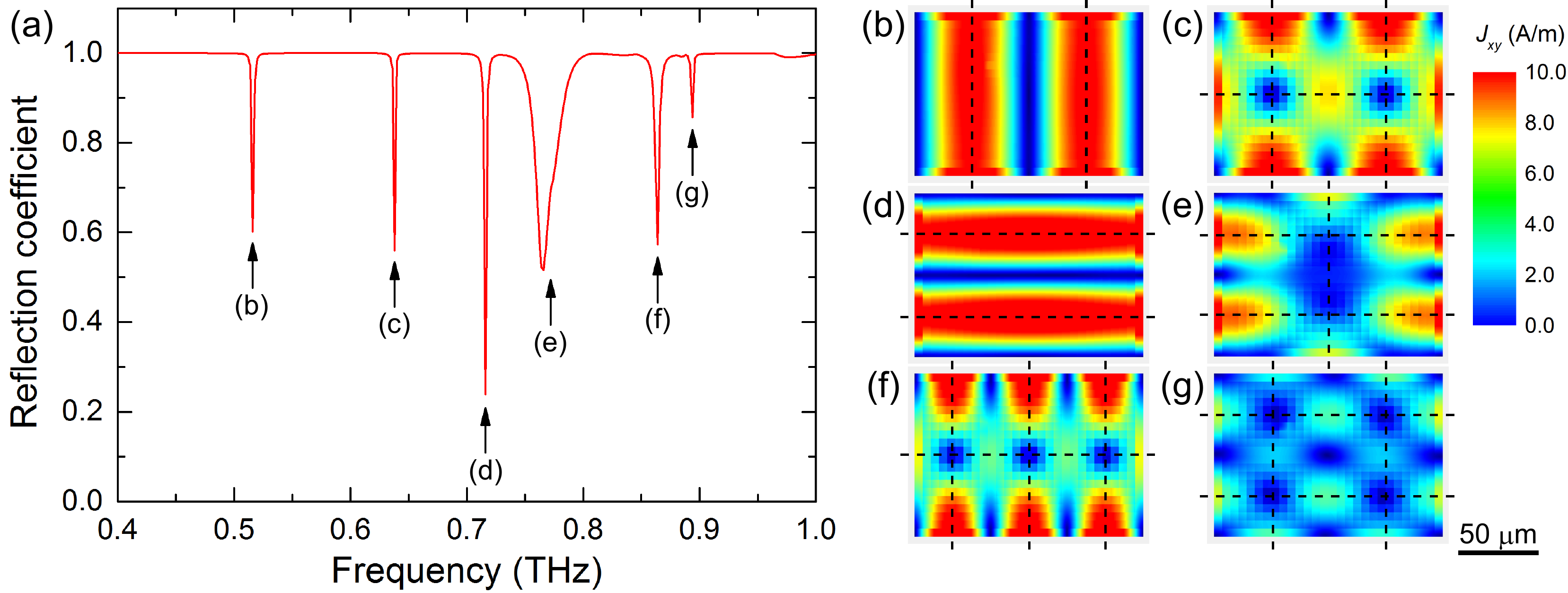} \\
	\caption{
(a) Simulation of the reflection coefficient of a 100$\times $140-$\mu $m$^2$ mesa, obtained from the scattering analysis.  
(b)--(g) Snapshots of the amplitude distribution of the in-plane surface current density $J_{xy}$ at the resonance frequencies.  Dashed lines in each panel represent positions of the line nodes of the EM standing wave.
}
\end{figure*}%

The simulated objects consist of the infinitely-thin superconducting patch atop the isotropic dielectric layer with thickness $t$.  We use the dielectric constant of $\epsilon =17.6$ from Ref.~\cite{Kadowaki10}.  The lateral sizes of the cell and the patch in the analytical space are 5$\times$5~$\mu$m$^2$ and 100$\times $140~$\mu $m$^2$, respectively. The patch is surrounded by electrical grounding 5 $\mu $m apart from the edges, which is consistent with the actual situation (see the image in the inset of Fig.~2(b)).  The EM feeding point with an area of $S=5$$\times$5~$\mu$m$^2$ is set near the center of the patch.  Note that the feed position has insignificant effect on the calculation result of the resonance frequency but affects the quality of the resonance according to the impedance matching.

Figure 5(a) shows the reflection coefficient at the feed point in the range of 0.4--1.0~THz.  Sharp drops are clearly observed at 0.516~THz, 0.638~THz, 0.716~THz, 0.766~THz, 0.864~THz, and 0.894~THz.  These drops are attributed to the EM emissions from the mesa to the free space.  The simulated resonance frequencies are consistent with $f_{mp}^c$ calculated from Eq.~(2).  In Fig.~5(a), the reflection coefficient is normalized by the characteristics impedance $Z$ at the feed point, which is determined in general by the ratio of the voltage to the current.  We assume $Z = v/J_c S$, where $v=V_{r}/N$ is the voltage per IJJ and $J_c=I_c /w \ell$ is the amplitude of the AC Josephson current.  Using realistic parameters $V_{r} = 1.0$~[V], $N=850$, and $I_c=18$~[mA], we obtain $Z =36.7$~$\Omega $.  In this particular case, we attribute the most intensive emission at the TM ($2,0$) mode to the impedance matching.

Snapshots of the amplitude distribution of the surface current density $J_{xy}$ at the resonance frequencies are displayed in Figs.~5(b)--(g).  The surface current is enhanced at the edges of the mesa owing to the edge singularity.  The simulated $J_{xy}$ is associated with the standing wave inside the mesa:  $J_{xy}=0$ and $\partial _{xy}J_{xy}=0$ correspond to the anti-node and the node, respectively.  The dashed lines in each panel represent line nodes of the standing wave.  We can identify the excited TM ($m,p$) mode from the numbers of the line nodes as follows:  (b) ($0,2$), (c) ($1,2$), (d) ($2,0$), (e) ($2,1$), (f) ($1,3$), and (g) ($2,2$).

\section{CONCLUSION}
The nearly square Bi-2212 mesa emits coherent and broadly tunable CW terahertz waves depending on the internal EM cavity resonance.  We used a collection optical system to directly identify the excited cavity modes.  The $f(V)$ that was simultaneously measured using a wedge-type interferometer verified the validity of the Josephson relation.  We analyzed the experimental data using the field simulator, and found that EM standing waves could form along the width and length of the mesa.  The most important new aspect of using a nearly square stack is that we can control the excited cavity mode and the concomitant polarization using an identical source.  Thus, compact and solid-state terahertz sources using the stacks of IJJs are thought suitable for coherent communication applications.

\begin{acknowledgments}
The authors thank K. Delfanazari, T. Kitamura, C. Watanabe, and H. Kambara for valuable discussions.  Bi-2212 single crystals were provided by T. Yamamoto.  Comments from H. Asai, R. Yoshizaki, and R. A. Klemm are greatly appreciated.  This work was supported by KAKENHI (Grants No. 13J04811, and No. 26790032).
\end{acknowledgments}
\newpage

\appendix

\section{Collimation effect of the silicon lens}
In order to collect the emitted integral terahertz waves from the sample, we used a collection optical system that consisted of a hemispherical silicon lens with a diameter of 5~mm, plastic lenses (PAX Co., Terahertz-Super Lens ``Tsurupica''), and off-axis parabolic mirrors.  Figures 6(a) and 6(b) show the far-field emission patterns that were measured separately using another system with and without the silicon lens, respectively.  In this case, the silicon lens was attached atop a conventional rectangular mesa with 60$\times $$300$~$\mu $m$^2$ that emits at the TM ($1,0$) mode.

The difference between the two cases clearly proves the collimation effect of the silicon lens that makes the pattern more unidirectional.  Based on simulations, the far-field pattern was found to be affected by the excited cavity mode~\cite{Hu09,Klemm11}.  The collecting angle of the plastic lens is more than 30$^{\circ}$; hence, the integral emission power from the sample can be monitored independently of the excited cavity mode.

\section{Wedge-type interferometer system}
In the measurement of the emission frequency $f$, we use a handmade wedge-type interferometer.  A wedge consisting of two quartz plates and a stainless-steel spacer was fixed on the 1D scanner, as shown in Fig.~1.  The raster scan was performed in the horizontal ($X$) direction, where the interspace distance of the wedge $d$ increases linearly with $X$.  The emitted monochromatic wave allows us to observe the interference pattern of the bolometer output.  As an example, we plot the bolometer output versus $X$ at 1.11~V in Fig.~7.  The solid line represents a sinusoidal wave function at the best fitted frequency from the least squares analysis.  Using this interferometer system, we can obtain $f$ with an uncertainty of 2\% much faster than using a commercial FT-IR spectrometer~\cite{Tsujimoto12-02}.

\begin{figure}[h] 
	\includegraphics[width=0.40\textwidth ,clip]{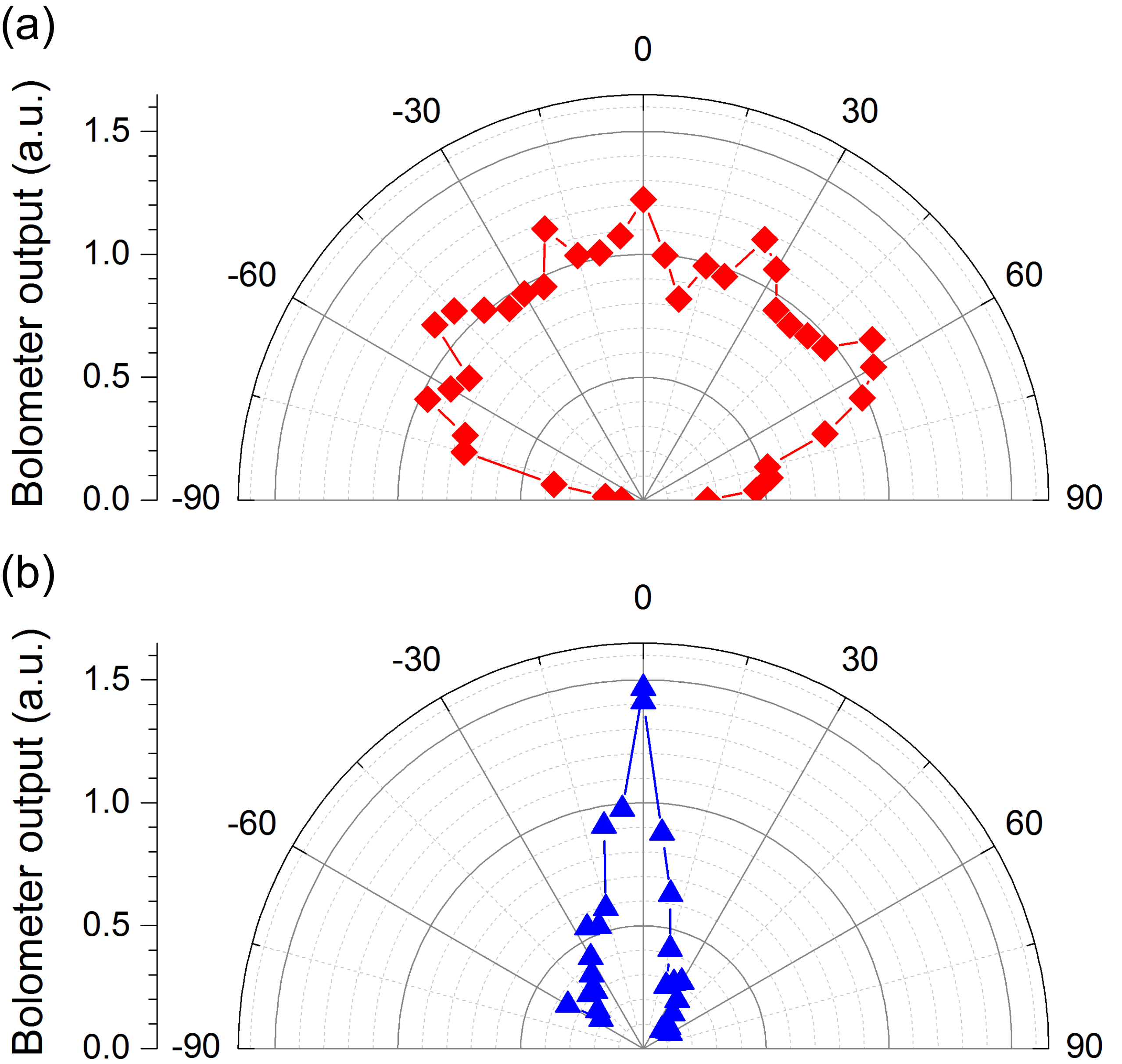} \\
	\caption{
Far-field emission pattern at the TM ($1,0$) mode measured (a) with and (b) without the collimation silicon lens. The detection angle varies in the E-field plane parallel to the mesa width.
}

\vskip15mm

	\includegraphics[width=0.40\textwidth ,clip]{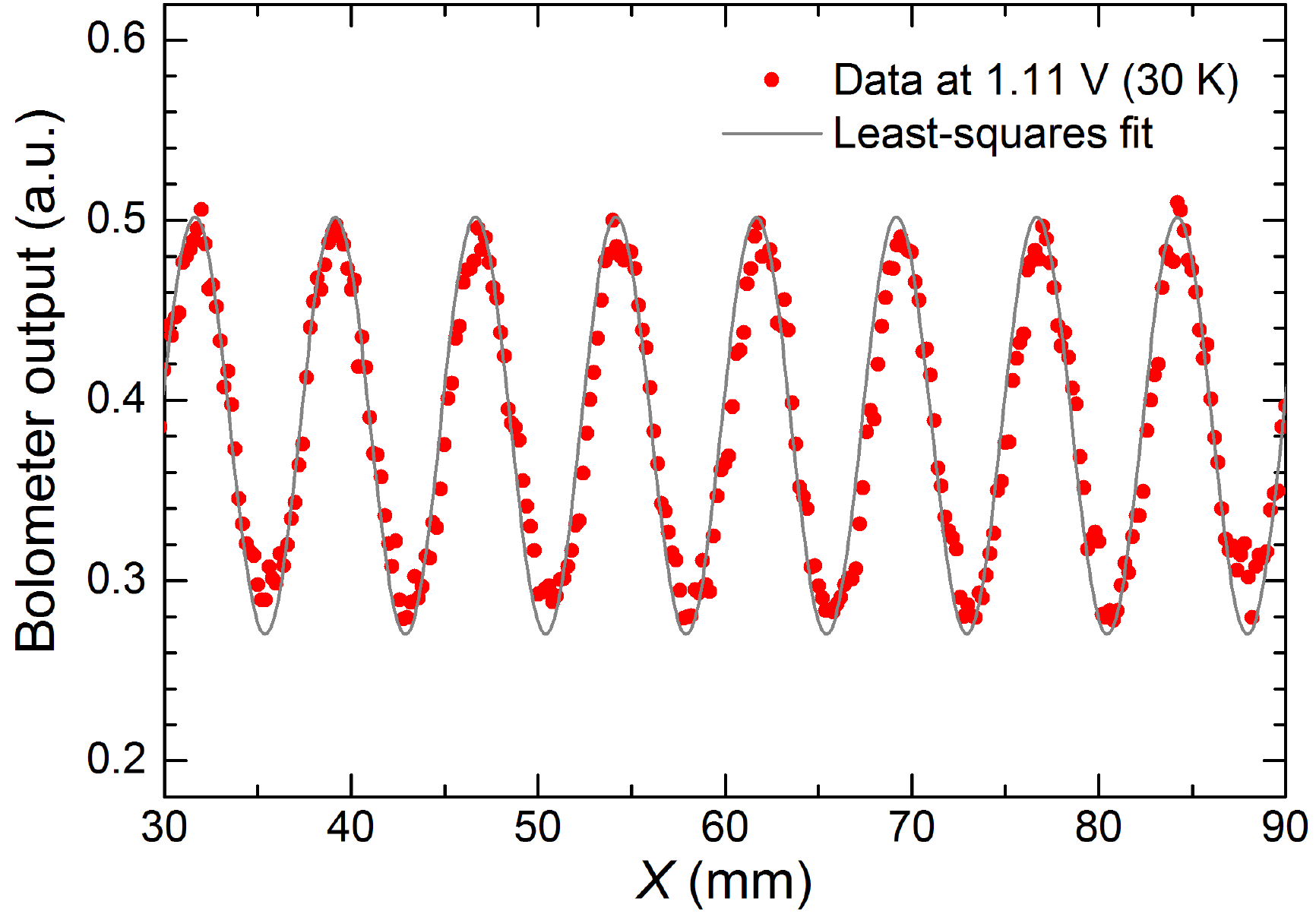} \\
	\caption{
Interference pattern obtained using the wedge-type interferometer.  The solid line represents a sinusoidal wave function at the best fitted frequency of 0.633~THz.
}
\end{figure}%


\newpage
\begin{thebibliography}{99}
\bibitem{Kawase04}  K. Kawase, Terahertz imaging for drug detection and large-scale integrated circuit inspection, Opt. Photonics News {\bf 15,} 34 (2004).
\bibitem{Tonouchi07} M. Tonouchi, Cutting-edge terahertz technology, Nature Photon. {\bf 1}, 97 (2007).
\bibitem{Kumar11} S. Kumar, Recent progress in terahertz quantum cascade lasers, IEEE J. Sel. Top. Quant. Elec. {\bf 17}, 38 (2011). 
\bibitem{Kanaya12} H. Kanaya, H. Shibayama, R. Sogabe, S. Suzuki, and M. Asada, Fundamental oscillation up to 1.31~THz in resonant tunneling diodes with thin well and barriers, Appl. Phys. Express {\bf 5}, 124101 (2012).
\bibitem{Welp13} U. Welp, K. Kadowaki, and R. Kleiner, Superconducting emitters of THz radiation, Nat. Photonics {\bf 7}, 702 (2013).
\bibitem{Kleiner92} R. Kleiner, F. Steinmeyer, G. Kunkel, and P. M\"{u}ller, Intrinsic Josephson effects in Bi$_2$Sr$_2$CaCu$_2$O$_8$ single crystals, Phys. Rev. Lett. {\bf 68}, 2394 (1992).
\bibitem{Josephson62}  B. D. Josephson, Possible new effects in superconductive tunnelling, Phys. Lett. {\bf 1}, 251 (1962).
\bibitem{Carver81}  K. R. Carver and J. W. Mink, Microstrip antenna technolog, IEEE Trans. Antennas Propag. {\bf 29}, 2 (1981).
\bibitem{Ozyuzer07} L. Ozyuzer, A. E. Koshelev, C. Kurter, N. Gopalsami, Q. Li, M. Tachiki, K. Kadowaki, T. Yamamoto, H. Minami, H. Yamaguchi, T. Tachiki, K. E. Gray, W.-K. Kwok, and U. Welp, Emission of coherent THz radiation from superconductors, Science {\bf 318}, 1291 (2007).
\bibitem{Kadowaki10} K. Kadowaki, M. Tsujimoto, K. Yamaki, T. Yamamoto, T. Kashiwagi, H. Minami, M. Tachiki, and R. A. Klemm, Evidence for a dual-source mechanism of terahertz radiation from rectangular mesas of single crystalline Bi$_2$Sr$_2$CaCu$_2$O$_{8+\delta }$ intrinsic Josephson junctions, J. Phys. Soc. Jpn. {\bf 79}, 023703 (2010).
\bibitem{Wang10} H. B. Wang, S. Gu$\acute{\rm{e}}$non, B. Gross, J. Yuan, Z. G. Jiang, Y. Y. Zhong, M. Gr\"{u}nzweig, A. Iishi, P. H. Wu, T. Hatano, D. Koelle, and R. Kleiner, Coherent terahertz emission of intrinsic Josephson junction stacks in the hot spot regime, Phys. Rev. Lett. {\bf 105}, 057002 (2010).
\bibitem{Tsujimoto12-01} M. Tsujimoto, T. Yamamoto, K. Delfanazari, R. Nakayama, T. Kitamura, M. Sawamura, T. Kashiwagi, H. Minami, M. Tachiki, K. Kadowaki, and R. A. Klemm, Broadly tunable subterahertz emission from internal branches of the current-voltage characteristics of superconducting Bi$_2$Sr$_2$CaCu$_2$O$_{8+\delta }$ single crystals, Phys. Rev. Lett. {\bf 108}, 107006 (2012).
\bibitem{Kitamura14} T. Kitamura, T. Kashiwagi, T. Yamamoto, M. Tsujimoto, C. Watanabe, K. Ishida, S. Sekimoto, K. Asanuma, T. Yasui, K. Nakade, Y. Shibano, Y. Saiwai, H. Minami, R. A. Klemm, and K. Kadowaki, Broadly tunable, high-power terahertz radiation up to 73K from a stand-alone Bi$_2$Sr$_2$CaCu$_2$O$_{8+\delta }$ mesa, Appl. Phys. Lett. {\bf 105}, 202603 (2014).
\bibitem{Kashiwagi11} T. Kashiwagi, K. Yamaki, M. Tsujimoto, K. Deguchi, N. Orita, T. Koike, R. Nakayama, H. Minami, T. Yamamoto, R. A. Klemm, M. Tachiki, and K. Kadowaki, Geometrical full-wavelength resonance mode generating terahertz waves from a single-crystalline Bi$_2$Sr$_2$CaCu$_2$O$_{8+\delta }$ rectangular mesa, J. Phys. Soc. Jpn. {\bf 80}, 094709 (2011).
\bibitem{Tsujimoto12-02} M. Tsujimoto, H. Minami, K. Delfanazari, M. Sawamura, R. Nakayama, T. Kitamura, T. Yamamoto, T. Kashiwagi, T. Hattori, and K. Kadowaki, Terahertz imaging system using high-$T_c$ superconducting oscillation devices, J. Appl. Phys. {\bf 111}, 123111 (2012).
\bibitem{Sonnet} See http://www.sonnetsoftware.com/ for details.
\bibitem{Li12} M. Li, J. Yuan, N. Kinev, J. Li, B. Gross, S. Gu$\acute{\rm{e}}$non, A. Ishii, K. Hirata, T. Hatano, D. Koelle, R. Kleiner, V. P. Koshelets, H. B. Wang, and P. Wu, Linewidth dependence of coherent terahertz emission from Bi$_2$Sr$_2$CaCu$_2$O$_8$ intrinsic Josephson junction stacks in the hot-spot regime, Phys. Rev. B {\bf 86}, 060505(R) (2012).
\bibitem{Kakeya13} I. Kakeya, N. Hirayama, T. Nakagawa, Y. Omukai, M. Suzuki, Temperature and current dependencies of terahertz emission from stacks of intrinsic Josephson junctions with thin electrodes revealed by a high-resolution FT-IR spectrometer, Physica C {\bf 491}, 11 (2013).
\bibitem{Tsujimoto14} M. Tsujimoto, H. Kambara, Y. Maeda, Y. Yoshioka, Y. Nakagawa, and I. Kakeya, Dynamic control of temperature distributions in stacks of intrinsic Josephson junctions in Bi$_2$Sr$_2$CaCu$_2$O$_{8+\delta }$ for intense terahertz radiation, Phys. Rev. Appl. {\bf 2}, 044016 (2014).

\bibitem{Hu09} X. Hu and S. Z. Lin, Cavity phenomena in mesas of cuprate high-$T_c$ superconductors under voltage bias, Phys. Rev. B {\bf 80}, 064516 (2009).
\bibitem{Klemm11} R. A. Klemm, E. R. LaBerge, D. R. Morley, T. Kashiwagi, M. Tsujimoto, and K. Kadowaki, Cavity mode waves during terahertz radiation from rectangular Bi$_2$Sr$_2$CaCu$_2$O$_{8+\delta }$ mesas, J. Phys.: Condens. Matter {\bf 23}, 025701 (2011).


\end{thebibliography}
\end{document}